\documentclass[%
 reprint,
superscriptaddress,
 amsmath,amssymb,
 aps,
]{revtex4-2}
 
\usepackage{graphicx}
\usepackage{dcolumn}
\usepackage{bm}
\usepackage{hyperref}

\usepackage{xcolor}

\begin{document}

\preprint{APS/123-QED}

\title{Ionization potential of radium monofluoride}
 
\author{S.~G.~Wilkins}
\email{wilkinss@mit.edu}
\affiliation{Massachusetts Institute of Technology, Cambridge, MA 02139}
\author{H.~A.~Perrett}
\email{holly.perrett@manchester.ac.uk}
\affiliation{School of Physics and Astronomy, The University of Manchester, Manchester, M13 9PL, United Kingdom}
\author{S.~M.~Udrescu}
\email{sudrescu@mit.edu}
\affiliation{Massachusetts Institute of Technology, Cambridge, MA 02139}
\author{A.~A.~Kyuberis}
\email{a.kiuberis@rug.nl}
\affiliation{Van Swinderen Institute for Particle Physics and Gravity, University of Groningen, Groningen, 9747~AG, The Netherlands}
\author{L.~F.~Pa\v{s}teka}
\affiliation{Van Swinderen Institute for Particle Physics and Gravity, University of Groningen, Groningen, 9747~AG, The Netherlands}
\affiliation{Department of Physical and Theoretical Chemistry, Faculty of Natural Sciences, Comenius University, Bratislava, 81806, Slovakia}
\author{M.~Au}
\affiliation{Systems Department,~CERN,~CH 1211~Geneva~23,~Switzerland}
\affiliation{Department Chemie, Johannes Gutenberg-Universität Mainz,~Mainz,~55099,~Germany}
\author{I.~Belo\u{s}evi\'{c}}
\affiliation{TRIUMF,~Vancouver,~BC ~V6T~2A3,~Canada}
\affiliation{Massachusetts Institute of Technology, Cambridge, MA 02139}
\author{R.~Berger}
\affiliation{Fachbereich~Chemie,~Philipps-Universit{\"a}t~Marburg,~Marburg,~35032,~Germany}
\author{C.~L.~Binnersley}
\affiliation{School of Physics and Astronomy, The University of Manchester, Manchester, M13 9PL, United Kingdom}
\author{M.~L.~Bissell}
\affiliation{School of Physics and Astronomy, The University of Manchester, Manchester, M13 9PL, United Kingdom}
\author{A.~Borschevsky}
\affiliation{Van Swinderen Institute for Particle Physics and Gravity, University of Groningen, Groningen, 9747~AG, The Netherlands}
\author{A.~A.~Breier}
\affiliation{Laboratory for Astrophysics,~Institute~of~Physics,~University~of~Kassel,~Kassel,~34132,~Germany}
\author{A.~J.~Brinson}
\affiliation{Massachusetts Institute of Technology, Cambridge, MA 02139}
\author{K.~Chrysalidis}
\affiliation{Systems Department,~CERN,~CH 1211~Geneva~23,~Switzerland}
\author{T.~E.~Cocolios}
\affiliation{KU Leuven,~Instituut~voor~Kern- en~Stralingsfysica,~Leuven,~B-3001 ,~Belgium}
\author{B.~S.~Cooper}
\affiliation{School of Physics and Astronomy, The University of Manchester, Manchester, M13 9PL, United Kingdom}
\author{R.~P.~de~Groote}
\affiliation{KU Leuven,~Instituut~voor~Kern- en~Stralingsfysica,~Leuven,~B-3001 ,~Belgium}
\author{A.~Dorne}
\affiliation{KU Leuven,~Instituut~voor~Kern- en~Stralingsfysica,~Leuven,~B-3001 ,~Belgium}
\author{E.~Eliav}
\affiliation{School of Chemistry, Tel Aviv University, Tel Aviv, 6997801, Israel}
\author{R.~W.~Field}
\affiliation{Massachusetts Institute of Technology, Cambridge, MA 02139}
\author{K.~T.~Flanagan}
\affiliation{School of Physics and Astronomy, The University of Manchester, Manchester, M13 9PL, United Kingdom}
\affiliation{Photon Science Institute,~The~University~of~Manchester,~Manchester,~M13~9PY,~United~Kingdom}
\author{S.~Franchoo}
\affiliation{Laboratoire Irène Joliot-Curie,~Orsay, F-91405, France}
\author{R.~F.~Garcia Ruiz}
\affiliation{Massachusetts Institute of Technology, Cambridge, MA 02139}
\author{K.~Gaul}
\affiliation{Fachbereich~Chemie,~Philipps-Universit{\"a}t~Marburg,~Marburg,~35032,~Germany}
\author{S.~Geldhof}
\affiliation{KU Leuven,~Instituut~voor~Kern- en~Stralingsfysica,~Leuven,~B-3001 ,~Belgium}
\author{T.~F.~Giesen}
\affiliation{Laboratory for Astrophysics,~Institute~of~Physics,~University~of~Kassel,~Kassel,~34132,~Germany}
\author{F.~P.~Gustafsson}
\affiliation{KU Leuven,~Instituut~voor~Kern- en~Stralingsfysica,~Leuven,~B-3001 ,~Belgium}
\author{D.~Hanstorp}
\affiliation{Department of Physics,~University~of~Gothenburg,~Gothenburg, 41296,~Sweden}
\author{R.~Heinke}
\affiliation{Systems Department,~CERN,~CH 1211~Geneva~23,~Switzerland}
\author{\'{A}.~Koszor\'{u}s}
\affiliation{Experimental Physics Department,~CERN,~CH 1211~Geneva~23,~Switzerland}
\author{S.~Kujanpää}
\affiliation{Department of Physics, Accelerator Laboratory, University of Jyväskylä, Jyväskylä, FI-40014, Finland}
\author{L.~Lalanne}
\affiliation{KU Leuven,~Instituut~voor~Kern- en~Stralingsfysica,~Leuven,~B-3001 ,~Belgium}
\author{G.~Neyens}
\affiliation{KU Leuven,~Instituut~voor~Kern- en~Stralingsfysica,~Leuven,~B-3001 ,~Belgium}
\author{M.~Nichols}
\affiliation{Department of Physics,~University~of~Gothenburg,~Gothenburg, 41296,~Sweden}
\author{J.~R.~Reilly}
\affiliation{School of Physics and Astronomy, The University of Manchester, Manchester, M13 9PL, United Kingdom}
\author{C.~M.~Ricketts}
\affiliation{School of Physics and Astronomy, The University of Manchester, Manchester, M13 9PL, United Kingdom}
\author{S.~Rothe}
\affiliation{Systems Department,~CERN,~CH 1211~Geneva~23,~Switzerland}
\author{A.~Sunaga}
\affiliation{Department of Physics, Graduate School of Science, Kyoto University, Kyoto 606-8502, Japan}
\affiliation{ELTE, E\"otv\"os Lor\'and University, Institute of Chemistry, P\'azm\'any P\'eter s\'et\'any 1/A 1117 Budapest, Hungary}
\author{B.~van~den~Borne}
\affiliation{KU Leuven,~Instituut~voor~Kern- en~Stralingsfysica,~Leuven,~B-3001 ,~Belgium}
\author{A.~R.~Vernon}
\affiliation{School of Physics and Astronomy, The University of Manchester, Manchester, M13 9PL, United Kingdom}
\author{Q.~Wang}
\affiliation{School of Nuclear Science and Technology,~Lanzhou~University,~Lanzhou,~730000, China}
\author{J.~Wessolek}
\affiliation{School of Physics and Astronomy, The University of Manchester, Manchester, M13 9PL, United Kingdom}
\author{F.~Wienholtz}
\affiliation{Experimental Physics Department,~CERN,~CH 1211~Geneva~23,~Switzerland}
\author{X.~F.~Yang}
\affiliation{School of Physics and State Key Laboratory of Nuclear Physics and Technology,~Peking~University,~Beijing,~100971,~China}
\author{Y.~Zhou}
\affiliation{Department of Physics and Astronomy, University of Nevada, Las Vegas, Las Vegas, Nevada 89154, USA}
\author{C.~Z\"{u}lch}
\affiliation{Fachbereich~Chemie,~Philipps-Universit{\"a}t~Marburg,~Marburg,~35032,~Germany}
\date{\today}

\begin{abstract}
The ionization potential (IP) of radium monofluoride (RaF) was measured to be 4.969(2)[10]~eV, revealing a relativistic enhancement in the series of alkaline earth monofluorides. The results are in agreement with a relativistic coupled-cluster prediction of 4.969[7]~eV, incorporating up to quantum electrodynamics corrections. Using the same computational methodology, an improved calculation for the dissociation energy ($D_{0}$) of 5.54[5]~eV is presented. This confirms that radium monofluoride joins the small group of diatomic molecules for which $D_{0}>\mathrm{IP}$, paving the way for precision control and interrogation of its Rydberg states.


\end{abstract}

\maketitle

The ionization potential (IP), defined as the minimum energy required to release a bound electron, is a fundamental property of atoms and molecules, and is important in chemistry and physics. 
The IP is central to revealing rigorous and intuitively appealing interrelationships among all electronic structural properties of atoms and molecules \cite{Ross2008}. 
Electronic states of these systems can be arranged into series, which differ in their principal quantum number, $n$, and excitation energy. 
The energy of states within each series increases with $n$ and eventually converges at the IP.

Due to the highly non-linear scaling of atomic and molecular properties with $n$, states with $n\gg1$ can exhibit remarkable properties when compared to their ground states.
These `Rydberg states' can have micrometer-sized atomic radii, extended lifetimes and extremely large transition dipole moments, thousands of times larger than those for their low-lying (low-$n$) states \cite{Gallagher1988}.
These exceptional characteristics make Rydberg states prominent systems for quantum computing and simulation \cite{Saffman2010}, precision measurements \cite{Jentschura2010} and the investigation of long-range interactions due to their high sensitivity to external electric fields and photon absorption.

For the majority of diatomic molecules, however, the dissociation energy ($D_{0}$) lies lower in energy than the IP. 
This severely limits the ability of experiments to study and manipulate their high-lying Rydberg states, as the molecules can fragment following pre-dissociation in short (ns) timescales.
A notable exception to this trend, with $D_{0}>\mathrm{IP}$, is BaF, which has been an important playground for the investigation of molecular Rydberg states owing to this rare property \cite{Zhou2015,Jakubek1995,Jakubek1994}.

Molecules containing isotopes of radium have been proposed as being promising systems in which to study the fundamental symmetries of the Universe \cite{Isaev2010,isaev2020lasercooled,Wilkins2023}, particularly in the hadronic sector of the Standard Model, where the rare octupole deformation of certain radium isotopes significantly boosts their sensitivity to $P,T$-violating phenomena \cite{Auerbach1996a}. A notable example is the RaF molecule, which was theoretically predicted \cite{Isaev2010} and later experimentally confirmed to be directly laser-coolable \cite{Udrescu2024,GarciaRuiz2020,AthanasakisKaklamanakis2024}. Leveraging the unique properties of high-lying Rydberg states of RaF through manipulating them with external fields, could enable the sensitive control of these molecules \cite{Hogan2012,Hogan2016}, offering complementary opportunities for future precision measurements. 
The realization of such techniques however relies upon the condition that $D_{0}>\mathrm{IP}$.

\begin{figure}[]
    \includegraphics[width=255pt]{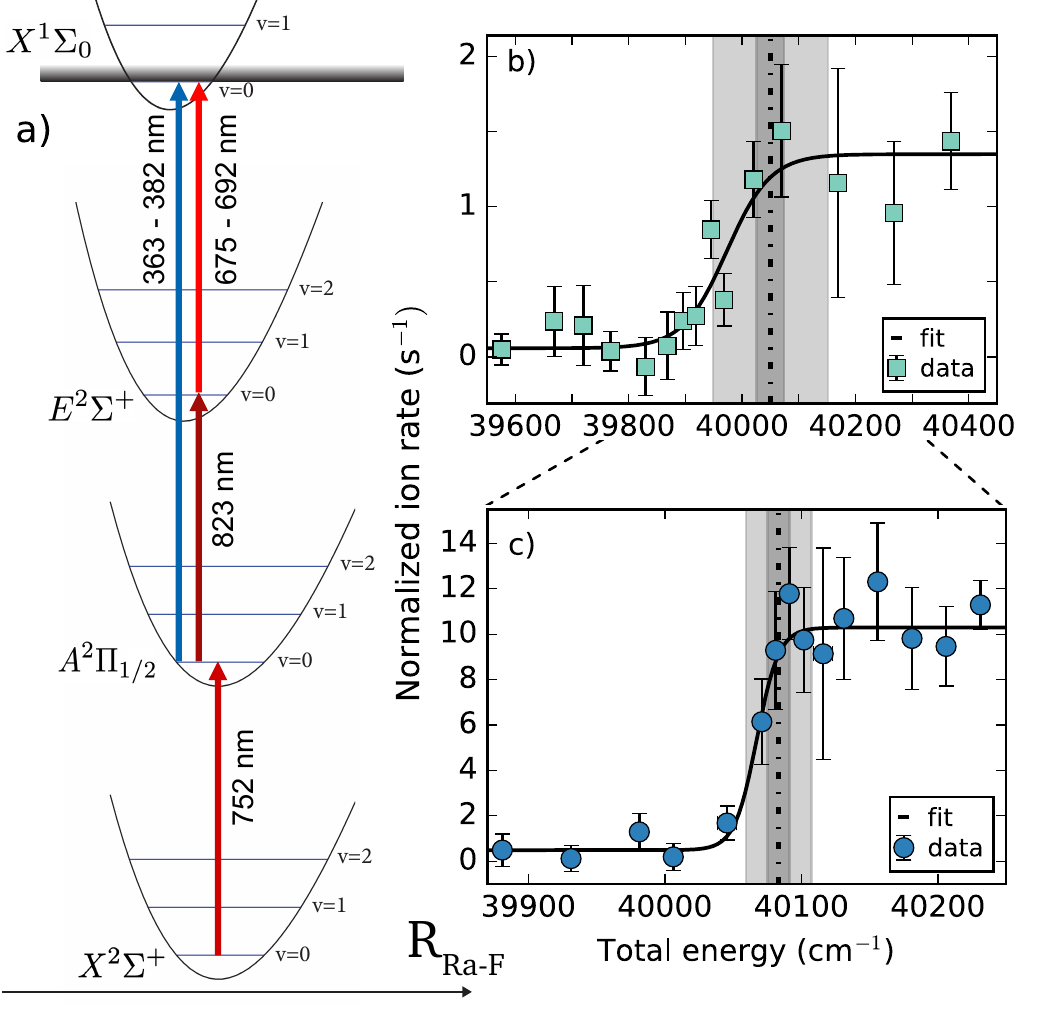}
    \caption{(Color online) \textbf{a)} The two-step and three-step ionization schemes used in the experiments. Power-normalized, background-subtracted ion count rate as a function of total photon energy in $^{226}$Ra$^{19}$F for the \textbf{b)} two-step scheme and \textbf{c)} three-step scheme. The determined IPs are shown as vertical dashed-and-dotted lines with their statistical and total uncertainties as dark and light gray bands, respectively.}
    \label{fig:RaF_IP}
\end{figure}

Here, we demonstrate that RaF possesses an exceptionally large dissociation energy that exceeds its ionization potential. 
We report the first measurement of the IP of $^{226}$Ra$^{19}$F, determined from the ionization threshold of its valence electron under multi-step laser excitation using both two-step and three-step ionization schemes. 
The obtained value 
is compared with \textit{ab initio} calculations performed within the relativistic coupled-cluster (RCC) framework, corrected for higher-order effects, including QED contributions. 
The same computational method is used to calculate its dissociation energy, resulting in an improved value, thus confirming that RaF possesses the rare property in which $D_{0}>\mathrm{IP}$.

\textit{Experimental details:}
Bunches of $^{226}$Ra$^{19}$F$^{+}$ were produced at the ISOLDE radioactive ion beam facility at CERN and studied with the Collinear Resonance Ionization Spectroscopy (CRIS) experiment. 
Details on the production of RaF molecules at ISOLDE can be found in Refs.~\cite{GarciaRuiz2020, Au2023}. 
Upon entering the CRIS beamline, the RaF$^{+}$ beam was neutralized in-flight after passing through a charge-exchange cell filled with a sodium vapor, where the neutral RaF molecules predominately populated the $X~^{2}\Sigma^{+}$ electronic ground state \cite{isaev2013ion}. 
The non-neutralized residual ions were deflected away, while the neutral bunches entered an ultra high-vacuum interaction region ($\approx~10^{-10}$~mbar), where they were collinearly overlapped with either two or three pulsed laser beams.

The ionization threshold was measured during two separate experiments. The ionization schemes for these are shown in Fig. \ref{fig:RaF_IP} a). In both experiments, the first laser excited the $A~^{2}\Pi_{1/2}(v=0) \leftarrow X~^{2}\Sigma^{+}(v=0)$ transition of the RaF molecules \cite{GarciaRuiz2020}, transferring molecules residing in multiple rotational states from the vibronic ground state \cite{Udrescu2024}. 
In the first experiment, the second laser was used to ionize molecules directly from the $A~^{2}\Pi_{1/2}(v=0)$ state, constituting a two-step resonance ionization scheme. 
In the second experiment, the discovery of the higher-lying $E~^{2}\Sigma^{+}$ state \cite{AthanasakisKaklamanakis2023} allowed a three-step resonance ionization scheme to be employed. This resulted in a significantly improved signal-to-background ratio as the superior pulse energy and beam quality of the ionization laser operating in its fundamental wavelength range increased the ionization efficiency while simultaneously decreasing the non-resonant background.
Additional details on the laser setups for each experiment can be found in the Supplemental Material.

The wavelengths of the ionization lasers were scanned and the resulting RaF$^{+}$ molecular ions were deflected onto an ion detector. The ion count rate was monitored as a function of ionization laser wavelength. The total excitation energy delivered to each molecule by the lasers was determined by Doppler-correcting the sum of the individual photon energies of the lasers in each ionization scheme.
Two scans at each ionization laser wavelength were taken; one in which all lasers of each scheme were present in the interaction region and the other one where only the ionization laser was present such that any background resulting from the ionization laser could be accounted for.
The ion rate was then determined as the difference between these two scans and linearly normalized with respect to the ionization laser power.

The resulting thresholds are shown in Fig. \ref{fig:RaF_IP} b) and c) for the two-step (teal squares) and three-step schemes (blue circles), respectively. 
Using the same method as presented in Ref.~\cite{Rothe2013}, the two sets of data were fit using a Sigmoid function and the IPs determined from the energy at which the ion rate saturates.
This was done by determining the energy at which the tangent to the inflection point of the fitted Sigmoid curves intersects the maximum ion rate, and its corresponding $\pm 1\sigma$ boundaries. The IPs and their statistical uncertainties are shown as the vertical dashed-and-dotted lines and the narrow dark grey bands in Fig. \ref{fig:RaF_IP} b) and c).
A systematic uncertainty from the fitting procedure is assigned by taking the energy difference between the inflection points of the curves and the extracted IPs. This is due to the lack of consistency in the literature regarding where the IP lies with respect to measured ionization thresholds. The total combined statistical and systematic uncertainties are shown as the wide light gray bands in Fig. \ref{fig:RaF_IP} b) and c).
More details on the fitting and its justification can be found in the Supplementary Material.
The ionization threshold measured using 
the three-step scheme (Fig. \ref{fig:RaF_IP} c)) can be seen to increase more sharply when compared to the two-step scheme (Fig. \ref{fig:RaF_IP} b)).
This is due to the additional rotational angular momentum ($J$) selectivity introduced by having two sequential resonant excitations before the ionization step in the three-step scheme.
This in turn reduced the number of high-lying Rydberg states populated by the ionization laser at energies below the IP, resulting in a sharper threshold. 
The ionization thresholds were determined to be 4.966(3)[10]~eV and 4.970(1)[2]~eV for the two-step and three-step scheme, respectively, with the statistical errors shown in curved brackets and the systematic errors shown in straight brackets.
More details on the fitting and its justification can be found in the Supplementary Material.

The majority of each molecular bunch was present in a shielded interaction region when the ionization process took place. 
However, a fraction of the molecules at the tail-end of each bunch were outside of this region where stray fields are non-negligible.
These molecules were gated out of the data set as a conservative measure to negate any systematic downward shift of the observed thresholds resulting from the electric fields present.
See the Supplementary Material for more details on the time-of-flight gating.

The potential energy curves for the $A~^{2}\Pi_{1/2}$, $E$~$^2 \Sigma ^{+}$ states in RaF and the $X~^{1}\Sigma^{+}$ RaF$^{+}$ ground state calculated here have similar equilibrium bond lengths computed to be 2.246~\r{A}, 2.191~\r{A} and 2.167~\r{A}, respectively.
As no significant change in the molecular geometry or vibrational quantum number occurs, the observed ionization thresholds from the $A~^{2}\Pi_{1/2}$ and $E$~$^2 \Sigma ^{+}$ state represent the adiabatic IP of RaF. 
The final experimental value is given as 4.969(2)[10]~eV, using the weighted standard deviation of the two-step and three-step scheme measurements to calculate the statistical error. The systematic errors from the two measurements were added in quadrature as the measurements were taken from separate independent experiments under different experimental conditions.

\textit{Computational method:} 
The relativistic single-reference coupled-cluster approach with single, double (CCSD), and perturbative triple excitations (CCSD(T)) was employed in our calculations, as implemented in the DIRAC19 program package \cite{Saue2020}. 
A zero-point energy (ZPE) correction of 5~meV is added to estimate the minimum vibrational energy of RaF and RaF$^{+}$ in the $^{2}\Sigma^{+}$ and $^{1}\Sigma^{+}$ states, respectively, using calculated harmonic vibrational frequencies. 
Relativistic core-valence-correlating Dyall basis sets~\cite{Dyall2016,dyall2009} of varying quality, cv$n$z ($n=2-4$) augmented by a single layer of diffuse functions, were used (s-aug-cv$n$z). 
The calculated potential energy curves were extrapolated to the complete basis-set limit (CBSL) using the scheme in Ref.~\cite{Feller1992} for the Dirac-Hartree-Fock energy and the scheme in Ref.~\cite{Helgaker1997} for the correlation contribution. In the calculations, 49 electrons were correlated and the virtual space was cut off at 50 a.u.

To correct for the limited active space used, the difference between results obtained correlating 49 electrons with a 50 a.u. cutoff and those obtained correlating all (97) electrons with a virtual space cutoff of 2000 a.u was calculated. In order to capture the full active space effect and to account for inner-core correlations, the all-electron quality basis set was used in the latter calculation. The more modestly sized dyall-cv3z and dyall-ae3z basis sets were employed, as calculations were prohibitively computationally expensive at the 4z level. Furthermore, the possible lack of diffuse functions was accounted for by taking  the difference between the d-aug-cv4z and the s-aug-cv4z results. The above corrections were calculated at a single geometry point and added to the potential energy curves.

The effect of perturbative triple excitations on the CBSL level was determined to be 51 meV. The effect of full triple excitations was evaluated to be around 1~meV by comparing the IP calculated at the CCSDT and CCSD(T) levels using the MRCC code \cite{Kallay2020,MRCC2}. These calculations were performed using the dyall.v3z basis sets, 16 correlated electrons, and a 10~a.u. virtual cutoff. Higher-level excitations were not considered owing to the very small difference between the CCSD(T) and CCSDT results.

\begin{table}[h!]
    \centering
    \caption{Experimental and calculated IP and $D_{0}$ of RaF.}\label{Tab:IPfinal}
    \begin{tabular}{lccc}
    \hline
Method & IP (eV) &$D_{0}$ (eV)\\
 \hline 
CBS-DC-CCSD&4.932\footnotemark[1]&5.427\\
CBS-DC-CCSD(T)&4.983\footnotemark[1]&5.520\\
+aug+ae.vs.cv & 4.977&5.538\\
+$\Delta$T & 4.978&5.534\\
+Breit & 4.976&5.532\\
+QED & 4.969&5.541\\
    \hline
Theoretical & 4.969[7]&5.54[5]\\
Experimental & 4.969(2)[10]&5.61(24)\footnotemark[2]\\ 
         \hline
    \end{tabular}
     \footnotetext[1]{ZPE correction is included.}
     \footnotetext[2]{Scaled from Ref. \cite{GarciaRuiz2020}.}
\end{table} 

In addition, the Breit and QED contributions were estimated. QED corrections were calculated using a development version of the DIRAC code~\cite{SunSalSau22}. Using this implementation, two effective QED potentials were added variationally to the DC Hamiltonian. The QED correction itself was obtained from a single-point calculation at the equilibrium geometry of neutral RaF. 
The Uehling potential \cite{Ueh35} was employed for vacuum polarization and the effective potential of Flambaum and Ginges for the electron self-energy \cite{FlaGin05}. Our estimate of the size of the Breit effect relied on the fact that the electronic structure of RaF is very similar to that of Ra$^+$ and upon ionization, the valence electron is removed from a Rydberg orbital (atomic-like and non-bonding). Thus, the contribution of the Breit interaction to the IP of RaF can be approximated by the effect calculated for the IP of Ra$^{+}$ (direct calculations of molecular Breit contributions are challenging at present). These calculations were performed within the Fock-space coupled-cluster approach (DCB-FSCC), using the Tel Aviv atomic computational package \cite{TRAFS-3C}.
 
The various higher-order corrections are added to the adiabatic CBS-DC-CCSD(T) IP and the final theoretical value was determined to be 4.969[7]~eV as shown in Table \ref{Tab:IPfinal}. Using the same computational method, the dissociation energy $D_{0}$ of RaF was calculated to be 5.54[5]~eV. The uncertainty on the calculated IP and $D_0$ was evaluated through further computational investigation using similar procedures that were previously employed for various properties of both atoms and molecules \cite{Hao2018,Guo2021,Haase2020,Leimbach2020}. The details of the uncertainty treatment for both properties can be found in the Supplemental Material. 



\textit{Discussion:}
Fig.~\ref{fig:RaF_IP_exp_theo_group_2s} a) shows experimental determinations of the IP of the group II monofluorides from CaF to RaF and compared with theoretical calculations using the same method as for RaF described previously \cite{Kyuberis2023}.
Experimental IP values measured using electron-impact \cite{Hildenbrand1965,Hildenbrand1968,Lau1989} and laser spectroscopy methods \cite{Berg1993,Jakubek1995} are shown as yellow diamonds and turquoise circles, respectively, with theoretical calculations shown as dark blue squares. 
The error bars for the laser measurements and theoretical predictions are smaller than the markers.
Fig.~\ref{fig:RaF_IP_exp_theo_group_2s} b) shows the deviation, in units of standard deviations, between the theoretical predictions (dark blue squares) and experimental values (gray line). Where they exist, the experimental values used in the comparison are from laser spectroscopy methods using Rydberg states or ionization threshold measurements, owing to their higher precision. In SrF, where no laser measurements exist, an electron-impact measurement is used for comparison. Excellent agreement within 1 standard deviation is obtained for CaF, BaF, and RaF whereas SrF agrees within 2 standard deviations.
Fig.~\ref{fig:RaF_IP_exp_theo_group_2s} a) highlights the ability to determine the IP of RaF at a comparable or better precision than experiments on the homologues, despite the technical challenges imposed by its short-lived nature.

The IP of the group II monofluorides decreases progressively with increasing proton number, before increasing again in RaF.
This enhancement in valence electron binding energy can be attributed to relativistic effects, which are significantly more pronounced in the heavier RaF. 
These effects result from the relativistic increase of the electron mass driven by the high velocity that bound electrons experience in atoms and molecules containing heavy elements \cite{Pyper2020}. 
This in turn spatially contracts orbitals with low angular momentum which increases their binding energies \cite{PyyDes79,Pyy88}.
The relativistic treatment required for correctly describing RaF is fully achieved in this study, yielding good agreement between theory and experiment as in the case of its homologues.

\begin{figure}[h]
    \includegraphics[width=250pt]{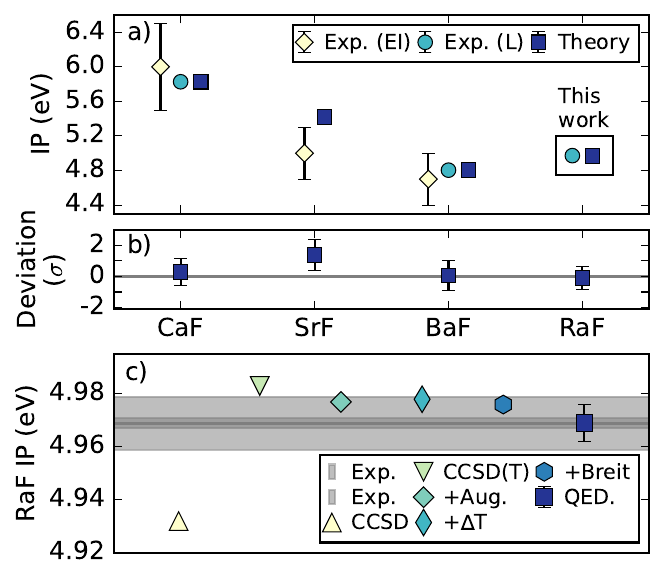}
    \caption{(Color online) \textbf{a):} Comparison of predicted theoretical and measured values for the IP of Group II monofluorides. Electron-impact (EI) \cite{Hildenbrand1965,Hildenbrand1968,Lau1989} and laser (L) measurements \cite{Berg1993,Jakubek1995} are denoted in the legend. The error bars for the laser measurements and theoretical predictions are smaller than the markers.
    \textbf{b):} 
    Deviation, in standard deviations, between the theoretical predictions (dark blue squares) and experimental values (gray line).
    \label{fig:RaF_IP_exp_theo_group_2s}
    \textbf{c):} 
    Comparison of experimental and theoretical values for the IP of $^{226}$RaF for increasing computational sophistication (left to right). An uncertainty is shown only for the most accurate theoretical value. The thin dark gray band and thick light gray band correspond to the statistical and systematic uncertainties on the experimental value for the RaF IP.}
    \label{fig:RaF_IP_exp_theo_group_2s}
\end{figure}

Fig.~\ref{fig:RaF_IP_exp_theo_group_2s} c) shows how the calculated IP for RaF evolves with an increasing level of computational sophistication. An agreement within 1 standard deviation is found between experiment and CCSD(T) theory that incorporates a complete basis set-extrapolation correction, non-perturbative triple excitations, and Breit and QED corrections. The final theoretical uncertainty is 7~meV, owing to the extensive implementation of higher-order contributions.

Benchmarking the predictive power of molecular theory is crucial for the study of radioactive molecules, which is a prominent future avenue for precision tests of the Standard Model~\cite{ArrowsmithKron2024}. 
As most of the short-lived radioactive molecules of interest can only be made in small quantities at specialized facilities, none of their chemical properties are known experimentally even though knowledge of their chemical behavior is necessary to guide, or even enable altogether, their production and study~\cite{Kirchner2003,Koester2008}.
Therefore, while the calculated IP of RaF is accurate even for predictions of lower sophistication in Fig.~\ref{fig:RaF_IP_exp_theo_group_2s}, the ability to reduce the theoretical uncertainty by including higher-order contributions, up to the QED level, remains important to aid experiments attempting to perform the first spectroscopy of previously unexplored systems. 

The previously calculated values of the ZPE-corrected dissociation energy, $D_{0}$, for RaF vary between 3.954~eV \cite{Isaev2010} and 5.255~eV \cite{isaev2013ion} with no corresponding uncertainties reported.
Using the same computational method as for the IP, which achieves a more extensive and rigorous treatment of higher-order contributions compared to previous studies, an improved calculation of the RaF $D_{0}$ equal to 5.54[5]~eV is presented. 
This is in agreement with the RaF $D_0$ extracted from Morse potential fits in Ref.~\cite{GarciaRuiz2020}, provided they are scaled by 1.55(6).  
Dissociation energy determinations from Morse potential fits use linear extrapolations that significantly underestimate the true dissociation energy in molecules where ionic bonding is important \cite{Gaydon1946,Beckel1971}.
By taking the extrapolated $D_{0}$ values for CaF~\cite{CharronCaF1995}, SrF~\cite{ColarussoSrF1996} and BaF~\cite{BernardBaF1992}, and comparing them to the true measured dissociation energies~\cite{JANAF1998}, the aforementioned scaling factor is obtained, which results in a RaF $D_{0}$ of 5.61(24)~eV (using the $X~^{2}\Sigma^{+}$ value from Ref.~\cite{GarciaRuiz2020}).
The resulting value and uncertainty of the computed $D_{0}$, corroborated by the agreement between experiment and theory for the IP, confirms that RaF joins BaF in the rare class of diatomic molecules for which $D_{0}>~$IP. 

 
\textit{Conclusion:}
The ionization potential of radium monofluoride (RaF) was measured with the CRIS experiment at CERN-ISOLDE to be 4.969(2)[10]~eV, through consistent observations of the ionization thresholds starting from two different excited electronic states.
The dissociation energy was found to lie above the IP, offering the possibility of accessing and utilizing Rydberg states in RaF, without loss due to pre-dissociation. This opens up several opportunities for controlling and manipulating the molecule via external fields \cite{Hogan2012,Hogan2016}. Additionally, these states can also offer an alternative window into probing the structure of radium nuclei \cite{Zhou2015}.


\textit{Acknowledgments:}
The authors thank Michael Morse for his insights. 
This work was supported by the Office of Nuclear Physics, U.S. Department of Energy, under grants DE-SC0021176 and DE-SC0021179,  the MISTI Global Seed Funds, Deutsche Forschungsgemeinschaft (DFG, German Research Foundation) -- Projektnummer 328961117 -- SFB 1319 ELCH; STFC grants ST/P004423/1 and ST/V001116/1, ERC Consolidator Grant no. 648381 (FNPMLS), Belgian Excellence of Science (EOS) project No. 40007501; KU Leuven C1 project No. C14/22/104, FWO project No. G081422N, International Research Infrastructures (IRI) project No. I001323N, the European Unions Grant Agreement 654002 (ENSAR2), LISA: European Union’s H2020 Framework Programme under grant agreement no. 861198, The Swedish Research Council (2016-03650 and 2020-03505), The National Key RD Program of China (No: 2022YFA1604800) (X.F.Y.) and the National Natural Science Foundation of China (No:12027809), the Dutch Research Council (NWO) projects number VI.C.212.016 and Vi.Vidi.192.088, the Slovak Research and Development Agency projects APVV-20-0098 and APVV-20-0127, JSPS Overseas Challenge Program for Young Researchers, Grant No. 201880193, JST Moonshot R\&D Grant No. JPMJMS2269 and NSF RII Track-4 (Grant No. 2327247). 
We thank the Center for Information Technology at the University of Groningen for their support and for providing access to the Peregrine and Hábrók high performance computing cluster.

\bibliography{apssamp}

\section{Supplemental Material}
\subsection{Experimental setup}
\textbf{Laser setup - two-step scheme:}
A pulsed grating-tunable Ti:Sa laser with a linewidth of several GHz \cite{Naubereit2014} produced 752-nm light used to drive the $A~^{2}\Pi_{1/2}(v=0) \leftarrow X~^{2}\Sigma^{+}(v=0)$ transition \cite{GarciaRuiz2020} of the RaF molecules, exciting multiple rotational states from the vibronic ground state. A maximum pulse energy of 150~$\mu$J was collimated through the interaction region with an 8~mm spot size.
The wavelength of this laser was kept constant and measured using a HighFinesse WSU-2 wavelength meter.

A Sirah Cobra pulsed dye laser with a linewidth of 1.8~GHz produced 726-764~nm light which was frequency doubled to 363-382~nm. This light was used to ionize molecules from the $A~^{2}\Pi_{1/2}(v=0)$ state. The pulse energy of the second harmonic output of this laser varied between between 50-130~$\mu$J and the light was collimated through the interaction region with a roughly 8-mm spot size.
The wavelength and power of the ionization laser were measured using a HighFinesse WS6 wavelength meter and Thorlabs S120VC power sensor, respectively.

The relative timings of the laser pulses were adjusted to maximize the above-threshold ion rate using a Quantum Composers 9520 Digital Delay Generator.

\textbf{Laser setup - three-step scheme:}
The 752-nm light used to drive the $A~^{2}\Pi_{1/2}(v=0) \leftarrow X~^{2}\Sigma^{+}(v=0)$ transition \cite{GarciaRuiz2020} of the RaF molecules was produced by a pulsed injection-seeded titanium-sapphire (Ti:Sa) laser \cite{Reponen2018} with a linewidth of around 20~MHz. A maximum pulse energy of 55~$\mu$J was delivered in a spot size of approximately 8~mm collimated through the interaction region. The combined high photon density and transition strength caused significant power broadening, intentionally exciting multiple rotational levels centered around $J=21.5$ from the vibronic ground state \cite{Udrescu2024}, despite the narrower linewidth of the laser. This was done to ensure the largest possible ion rate from which to determine the ionization threshold, given the challenges in studying such a small molecular beam intensity (10$^{5}$~s$^{-1}$).

A pulsed grating-tunable Ti:Sa laser with a linewidth of a several GHz \cite{Naubereit2014} produced 823-nm light to further excite the molecules residing in $A~^{2}\Pi_{1/2}(v=0)$ to the $E~^{2}\Sigma^{+}(v=0)$ state. A maximum pulse energy of 130~$\mu$J was collimated through the interaction region with an 8~mm spot size.
The wavelengths of the lasers used for the first two excitation steps were kept constant and measured using a HighFinesse WSU-2 wavelength meter.

A Sirah Cobra pulsed dye laser with a linewidth of 1.8~GHz produced 675-692~nm light to ionize molecules from the $E~^{2}\Sigma^{+}(v=0)$ state. The pulse energy of this laser varied between 780-900~$\mu$J across this wavelength range and the light was collimated through the interaction region with an 8-mm spot size. 
The wavelength and power of the ionization laser were measured using a HighFinesse WS6 wavelength meter and Thorlabs S370C power sensor, respectively.

The relative timings of the laser pulses were adjusted to maximize the above-threshold ion rate using a Quantum Composers 9520 Digital Delay Generator.

\subsection{Data taking, processing and analysis}
\textbf{Doppler correction:} The total excitation energy delivered to the molecules was determined by first adding the photon energies of the lasers in each scheme, as read by the two wavelength meters. This sum, $E_{\mathrm{Lab}}$, was converted to the molecular rest frame energy,  $E$, according to:
\begin{equation}
    E=E_{\mathrm{Lab}}\frac{\sqrt{1-\beta^{2}}}{(1+\beta)},
\end{equation}
with
\begin{equation}
    \beta = \sqrt{1+\frac{m^{2}c^{4}}{(eV+mc^{2})^{2}}},
\end{equation}
where $m$ is the combined masses of $^{226}$Ra and $^{19}$F \cite{Huang2021}, 
$e$ is the charge of the electron and $c$ is the speed of light and $V$ is the acceleration voltage which averaged 39954(1)~V for the two-step scheme experiment and 29614(1)~V for the three-step scheme experiment.
The uncertainty on the Doppler-corrected total photon energy, $E$, was determined to be much smaller than 1~cm$^{-1}$.


\textbf{Scan procedure:} Data was taken where the ion count rate was measured at discrete ionization laser wavelengths across the threshold region.
Two scans at each set wavelength were taken after a measurement of the ionization laser power. In the first, all lasers for each ionization scheme were present in the interaction region and in the second, only the ionization lasers were present. This allowed any background resulting from the ionization laser to be accounted for.
The ion rate was then determined as the difference between these two scans and was normalized with respect to the ionization laser power, which was measured prior to each scan.
The data was then binned with respect to energy.

\begin{figure}
    \centering
    \includegraphics[width=\columnwidth]{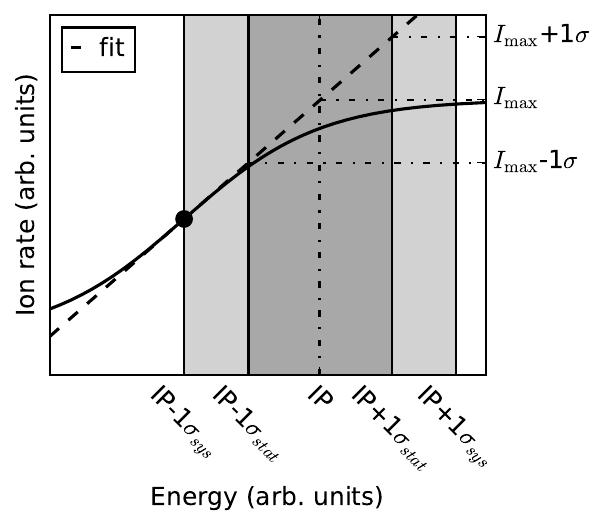}
    \caption{Schematic representation of IP fitting procedure. See text for details.}
    \label{fig:IPfit}
\end{figure}

\textbf{Fitting procedure:} There is a lack of consistency in the literature on how to extract the ionization potential (IP) from observed ionization thresholds. 
Here, the same procedure from Ref. \cite{Rothe2013} was followed where the IP is determined from the energy at which the ion rate saturates. 
This is justified in Ref. \cite{Rothe2012}, which contains an extended discussion of IP measurements derived from ionization thresholds obtained in resonance ionization experiments.

This method was chosen, instead using the onset energy where the first ions are observed or inflection point of the ion rate curve, because high-lying Rydberg states are expected to be populated by the laser ionization scheme in the experiment.
These states can be subsequently ionized non-resonantly by absorption of another photon from the ionization laser or by the electric field generated by the deflection plates at the end of the interaction region with latter mechanism only being possible for states which lie sufficiently close in energy to the IP. 
The high-lying Rydberg states appear in series which increase in density as the IP is approached, forming a threshold. 
Once a certain energy is reached, all nearby photon energies result in a consistent ion rate, resulting from directly ionizing molecules from the excited states with a single photon, which is interpreted as the IP.

The binned data was fitted using a Sigmoid function given by the following: 
\begin{equation}
    I(E) = B + \frac{I_{\mathrm{max}}-B}{1+e^{(E_{0}-E)/k}}
\end{equation}
where $I(E)$ is the energy-dependent $^{226}$RaF ion rate, $B$ is the background, $I_{\mathrm{max}}$ is the maximum ion rate, $k$ is the threshold width and $E_{0}$ is energy of the inflection point of the curve.

A schematic representation of the IP fitting procedure is shown in Fig.~\ref{fig:IPfit}. 
The best-fit Sigmoid function is shown as the curved solid black line and its inflection point as a solid black circle. 
The tangent to the inflection point of the best-fit Sigmoid function is shown as the diagonal dashed black line, which intersects ion rates corresponding to $I_{\mathrm{max}}-1\sigma$, $I_{\mathrm{max}}$ and  $I_{\mathrm{max}}+1\sigma$.
The IP, shown as the vertical dashed-and-dotted line, is determined as the energy at which the tangent of the best-fit Sigmoid curve intersects $I_{\mathrm{max}}$, with its statistical uncertainty being determined by the energies at which it intersects $I_{\mathrm{max}}\pm 1\sigma$.
The statistical uncertainty on the IP is shown as a vertical dark gray band.
The systematic uncertainty from the fitting procedure, resulting from the inconsistency in the literature, is assigned by taking the energy difference between the inflection point of the curves and the extracted IP.
This systematic uncertainty is shown as a vertical light grey band in Fig.~\ref{fig:IPfit}.

\textbf{Stray electric field impact:}
Electric fields present during ionization are known to decrease the observed IP according to $\Delta\mathrm{(IP)~(cm}^{-1})=6.12\sqrt{\epsilon}$  
where $\epsilon$ is the electric field in V/cm \cite{Linton1999,Chupka1993}.
The laser excitation and ionization process during both experiments predominately took place in a shielded interaction region.  
Outside of the shielded region are electrostatic elements held at high voltage including the non-neutral beam deflector (2000~V) on one side and the post-ionization deflection plates ($\pm 2600$~V) on the other. 

\begin{figure}
    \centering
    \includegraphics[width=\columnwidth]{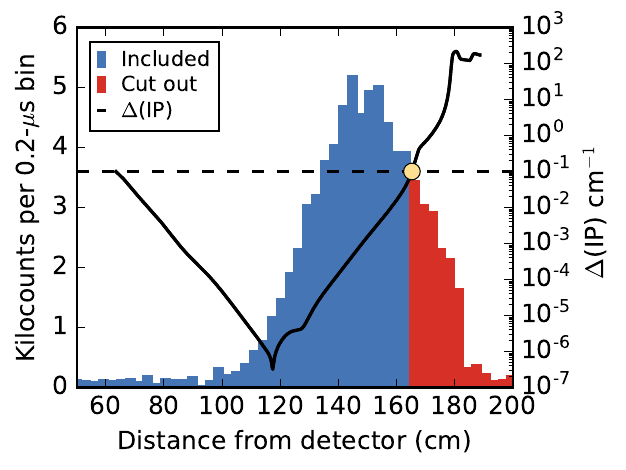}
    \caption{$\Delta\mathrm{(IP)}$ calculated as a function of distance from the ion detector (black line). The horizontal dashed line corresponds to the limit set at $\Delta\mathrm{(IP)}=0.1~\mathrm{cm}^{-1}$. Time-of-flight profile of $^{226}$RaF$^{+}$ ions in the three-step scheme experiment is shown as a histogram. The blue portion of the histogram was included in the data analysis whereas the red portion was excluded.}
    \label{fig:DeltaIPEfield}
\end{figure}

To estimate the magnitude of any fields that leak into the shielded region, SIMION \cite{simion} simulations were performed and the quantity $6.12\sqrt{\epsilon}$ was evaluated across the central axis of the vacuum chambers, in addition to 8 equidistant lines that follow the surface of a cylinder with a radius of 1~cm surrounding this. The simulated central-axis values are shown in Fig. \ref{fig:DeltaIPEfield} as a black line. 
From this, it is shown that the tail-end of the $^{226}$RaF$^{+}$ bunches in each experiment were in a region with non-negligible electric fields.
Within the shielded region, the simulated off-axis $\Delta\mathrm{(IP)}$ values were in agreement with the on-axis value. However, outside of this region, the off-axis values varied significantly.
As the exact beam profile and propagation axis were not known during the experiments, cutting the data with a conservative limit of $\Delta\mathrm{(IP)~(cm}^{-1})>0.1$ was chosen in favor of applying a energy correction as a function $^{226}$RaF$^{+}$ ion time of flight.


\subsection{Computational method details}


The final calculated IP and $D_0$ of RaF are shown in Table \ref{Tab:IPfinal}, where the size of different contributions to the final values is also presented.
The zero-point energy (ZPE) correction for RaF and
RaF$^{+}$ in the $^{2}\Sigma^{+}$ and $^{1}\Sigma^{+}$ states was based on the calculated vibrational frequencies of 438~cm$^{-1}$ and 518~cm$^{-1}$, respectively.

The main sources of uncertainty in the calculations are the incompleteness of the employed basis set, the approximations in treating electron correlations and the missing higher-order relativistic effects. These are assumed to be largely independent of each other, and hence are investigated separately.

\begin{table}[h!]
    \centering
    \caption{Experimental and calculated IP and $D_{0}$  of RaF.}\label{Tab:IPfinal}
    \begin{tabular}{lccc}
    \hline
Method & IP (eV) &$D_{0} (eV)$\\
 \hline 
CBS-DC-CCSD&4.932\footnotemark[1]&5.427\\
CBS-DC-CCSD(T)&4.983\footnotemark[1]&5.520\\
+aug+ae.vs.cv & 4.977&5.538\\
+$\Delta$T & 4.978&5.534\\
+Breit & 4.976&5.532\\
+QED & 4.969&5.541\\
    \hline
Theoretical & 4.969[7]&5.54[5]\\
Experimental & 4.969(2)[10]&5.61(24) \\
         \hline
    \end{tabular}
      \footnotetext[1]{ZPE correction is included.}
\end{table}


\textbf{Basis set - cardinality}: The final results were obtained through a complete basis-set limit (CBSL) extrapolation of the s-aug-cv$n$z basis sets following the H-CBSL scheme \cite{Helgaker1997} for the correlation contribution. While this is a popular CBSL extrapolation approach, two additional schemes were tested; the Martin $(n+\frac{1}{2})^{-4}$ scheme \cite{Martin1996} (M) and the scheme of Lesiuk and Jeziorski \cite{Lesiuk2019} (LJ). 


The results obtained using the M and LJ schemes are consistent to within 1 meV with respect to the H-CBSL value, confirming the convergence of the calculated IP with respect to the basis-set cardinality. The uncertainty due to the incompleteness of the basis set is therefore estimated to be half the difference between the H-CBSL and s-aug-cv4z values.

\textbf{Basis set - augmentation:} The uncertainty due to the possible insufficient number of diffuse functions is evaluated as the difference between the results obtained using the doubly augmented and the singly augmented dyall.cv4z basis sets, see Table~\ref{Tab:uncertainty}.



\textbf{Correlations - core:} To account for the uncertainty due to incomplete size of the active space and any shortcomings in the treatment of core correlations, half the size of corrections applied in Tables~\ref{Tab:IPfinal}.


\textbf{Correlations - Higher-order excitations:} The contributions of $\Delta$T in Tables~\ref{Tab:IPfinal} to estimate the uncertainty due to the missing higher-order excitations.


\textbf{Relativity - QED effects:} Higher-order QED contributions are assumed to be smaller than the calculated Lamb shift which is given as an conservative uncertainty estimate \cite{Kyuberis2023}. 


 \begin{table}
         \centering
         \caption{Summary of the main sources of uncertainty in the calculated IP and $D_{0}$ of RaF.}\label{Tab:uncertainty}
         \begin{tabular}{llcc}
         \hline\hline
               Category & Error source & IP (meV) &$D_{0}$ (meV) \\
               \hline
               Basis set & Cardinality & 2.6 & 50.1 \\
                         & Augmentation& 0.3 &3.3\\
               Correlation& Core electrons & 4.4 &4.9\\ 
                         & Higher-order excitations &0.9&4.0\\
               Relativity&  QED & 4.2 &5.7 \\
         \hline 
         Uncertainty: && 7 & 51\\
         \hline
         \end{tabular}
     \end{table} 

The magnitude of the various effects contributing to the total uncertainty on the calculated IP and $D_0$ is given in Tables~\ref{Tab:uncertainty}. The dissociation energy is determined by the strength of the bonding, which is more sensitive to the basis set quality than the ionization potential, leading to a higher uncertainty. The total uncertainty is obtained by taking the Euclidean norm of the individual uncertainties, as these are assumed to be independent.  
\end{document}